\documentclass{jpsj3}
\usepackage{txfonts}
\usepackage{graphicx}  
\usepackage{amsmath,amssymb,bm}
\usepackage{url,cite}       
\usepackage{here} 
\usepackage{booktabs}
\usepackage{multirow}
\usepackage{color}

\title{Performance of the Extended Ising Machine for the Quadratic Knapsack Problem}

\author{Haruka Akishima$^1$\thanks{Present address: Global R\&D Center for Business by Quantum-AI Technology, National Institute of Advanced
Industrial Science and Technology (AIST), Ibaraki, Japan, akishima.haruka@aist.go.jp}, Hirotaka Tamura$^2$, and Kazue Kudo$^{1,3}$\thanks{kudo@is.ocha.ac.jp}}
\inst{$^1$Department of Computer Science, Ochanomizu University, Bunkyo, Tokyo 112-8610, Japan \\
$^2$DXR Laboratory, Inc., Yokohama 223-0066, Japan \\
$^3$Graduate School of Information Sciences, Tohoku University, Sendai 980-8579, Japan} 

\abst{
    {
        The extended Ising machine (EIM) enhances conventional Ising models, which handle only binary quadratic forms by allowing constraints through real-valued dependent variables. We address the quadratic knapsack problem (QKP), hard to solve using Ising machines when formulated as a quadratic unconstrained binary optimization (QUBO). We demonstrated the EIM's superiority by comparing it with the conventional Ising model-based approach, a commercial exact solver, and a state-of-the-art heuristic solver for QKP.
    }
}

\begin{document}
\maketitle


The Quadratic Knapsack Problem (QKP) is a combinatorial optimization problem that maximizes quadratic profit under linear constraints. QKP has various applications, including optimizing the allocation of communication facilities and planning transportation hub locations\cite{pisinger2007quadratic}. However, owing to its quadratic nature and NP hardness, QKP remains difficult to solve.

The Ising machine is a computational system specialized in combinatorial optimization problems. After the materialization of quantum-annealing theories into a quantum-annealing machine\cite{johnson2011quantum}, 
 there has been active development of classical Ising machines implemented in classical devices\cite{aramon2019physics 
 }. To solve a combinatorial optimization problem using an Ising machine, the problem must be mapped to an Ising model or quadratic unconstrained binary optimization (QUBO). { When converting QKP constraints into QUBO formulations, it is often difficult to solve small instances\cite{jimbo2022hybrid, 
parizy2021analysis, ohno2024toward}. We demonstrate that this difficulty can be overcome by using an extended Ising machine (EIM) that adds dependent variables representing various constraints to the conventional Ising model.}

The EIM \cite{yin2023extended,watanabe2024nebula} is an extension of the conventional Ising model that efficiently handles inequality constraints and higher-order cost functions. Using binary variables $x_i \in \{0,1\}$, the objective function, or Hamiltonian, that the EIM minimizes is expressed as
\begin{equation}
H(\bm{x}) = -\frac{1}{2} \sum_{i=1}^n \sum_{j=1}^n W_{i,j}x_ix_j - \sum_{i=1}^n b_ix_i + \sum_{k=1}^m \lambda_k G_k(r_k),\\
\label{eq:H_extended}
\end{equation}
where $n$ is the number of decision variables, $m$ is the number of constraints, $W_{i,j}\ (W_{i,j}=W_{j,i}, W_{i,i}=0)$ is the interaction coefficient between the variables $x_i$ and $x_j$, and $b_i$ is the coefficient of the linear terms. The first and second terms on the right side correspond to those of the QUBO formulation. The EIM includes the sum of the real-valued dependent variables $G_k(r_k)$ in the third term. 
Function $G_k$ represents inequality constraints or higher-order terms, taking the resource variable $r_k$ as input. The resource variable $r_k$ is a linear sum of the variables $x_i$, depending on the coefficient $Z_{k,i}$ and the constant $c_k$, and is expressed as
\begin{equation}
    r_k = \sum_{i=1}^n Z_{k,i}x_i + c_k.
\end{equation}
The coefficient $\lambda_{k}$ in Eq. \eqref{eq:H_extended} controls the influence of the $k$-th dependent variable and is positive for inequality constraints. 

The inequality constraints are represented using the dependent variable, expressed as
\begin{equation}
    G_k(r_k) = \max(0, r_k).
    \label{eq:G_k_ineq}
\end{equation}
Note that $G_k(r_k) = 0$ if constraint $k$ is satisfied and $G_k(r_k)=r_k$ otherwise.

The architecture of EIM is an extension of Digital Annealer\cite{aramon2019physics,yin2023extended}, the classical Ising machine. The optimization algorithm combines the Markov chain Monte Carlo (MCMC) with the rejection-free selection rule\cite{bortz1975new,rosenthal2021jump} and the replica-exchange Monte Carlo (RMC) method\cite{PhysRevLett.57.2607,hukushima1996exchange}.

The QKP considered in this study is formulated as follows:
\begin{equation}
    \begin{aligned}
        & \text{maximize}
            & \sum_{i=1}^{n} \sum_{j=i}^{n} p_{i,j} x_i x_j,\\
        & \text{subject to}
            & \sum_{i=1}^{n} w_{i}x_i \le c,\\
    \end{aligned}
\end{equation}
where $n$ is the number of items, $w_i$ is the weight of item $i$, $c$ is the knapsack capacity, $p_{i, j}$ is the profit of the item pairs $i$ and $j$, and $p_{i,i}$ is the profit of item $i$. The binary variable $x_i = 1$ if item $i$ is selected and $x_i = 0$ otherwise.

We conducted numerical experiments using the benchmark set of the QKP\cite{
    billionnet2004using, BILLIONNET2004565}. The number of items $n$ is 100, 200, or 300. The profits \( p_{i,j} \) and weights \( w_i \) are uniformly distributed integers in ranges 1--100 and 1--50, respectively, whereas the capacity \( c \) is a randomly selected integer between 50 and $\max(50, \sum_{i=1}^{n} w_i)$. 
{
The density of the profit matrix $p_{i,j}$ for each instance—i.e., the ratio of nonzero elements in the upper triangular part of the matrix (\(i \le j\))—was 25\%, 50\%, 75\%, and 100\% for instances with $n = 100$ and $n = 200$,  and 25\% and 50\% for instances with $n = 300$. We used ten randomly generated instances for each of these ten \((n, \text{density})\) combinations, resulting in 100 instances.}

{Since the exact optimal solutions for these instances are known\cite{parizy2021analysis,ohno2024toward}, the performance of each method was evaluated according to whether it could reach the optimal solution.}

{
Numerical simulations demonstrate that solving QKP is easier when the constraints are formulated directly without converting them to QUBO.} The hardware used was an Intel Core i7-13700 (16 core, 2.1 GHz) with 64 GB DDR5 RAM.

For the numerical simulations, the cost function $H_{\rm{cost}}$, which represents the profit of QKP, is common to the QUBO and EIM formulations. It is expressed as
\begin{equation}
H_{\rm{cost}} = -\sum_{i=1}^{n} \sum_{j=1}^{n} p_{i,j} x_i x_j.
\end{equation}
In the EIM formulation, the problem is expressed as
\begin{equation}
H_{\rm{qkp\_extend}} = H_{\rm{cost}} + \lambda_{\rm{extend}} \max\left(0, \sum_{i=1}^{n} w_i x_i - c\right),
\label{eq:qkp_eim}
\end{equation}
where $n$ binary variables and a dependent variable are required.
For comparison, the QUBO was formulated as follows.
\begin{equation}
    H_{\rm{qkp\_qubo}} = H_{\rm{cost}} + \lambda_{\rm{qubo}} \left(\sum_{i=1}^{n}w_ix_i + \sum_{j=0}^{\lfloor \log_2 c \rfloor}2^jy_j - c\right)^2,
    \label{eq:qkp_qubo}
  \end{equation}
where $y_{j} \in \{0,1\}$ and $\lfloor \cdot \rfloor$ denote floor functions. 
The required number of binary variables is $n + \lfloor \log_2 c \rfloor + 1$.

The penalty coefficients $\lambda_{\rm{qubo}}$ and $\lambda_{\rm{extend}}$ were selected by a grid search with a relative accuracy of $0.1$ to minimize the time for reaching a constraint satisfaction solution.

This section describes the temperature settings of the RMC used in the experiment. 
The number and arrangement of the replicas were set based on previous studies \cite{kone_selection_2005, atchade2011towards} which suggested that an exchange probability of 23\% between adjacent replicas is optimal. 
By following the approach described in Ref. \citen{hukushima1999domain}, we conducted a short RMC simulation to determine the temperatures such that the average exchange probability between each adjacent replicas is around 20\%.
Temperature-exchange trials were performed every 10 MCMC runs, alternating between even and odd replica pairs.
The highest temperature was set to a value that would result in search behavior equivalent to random sampling. To achieve this goal, the temperature was selected such that the energy variance obtained from $10^5$ MCMC steps was equivalent to that obtained from random sampling. The lowest temperature was determined by running $10^5$ RMC simulations using the highest temperature and arrangement described above so that state transitions would occur with a reasonable frequency at the lowest temperature. As a measure of the frequency of state transitions, we used the proportion of the most frequent state in the histogram of the sampled states and found that an efficient search was performed when this state made up approximately 10\% of the total histogram at the lowest temperature.

\begin{table*}[t]
    \centering
    \caption{\label{tab:result}
    Performance comparison. \#Optimal refers to the number of instances that achieved the optimal solution with at least one random seed. 
    The average run time indicates the average computational time calculated only over the instances (and seeds, for the EIM and QUBO) that reached the optimal solution.
    The measurement results of IHEA and Gurobi are published in Ref. \citen{hochbaum2025fast}. Note that the execution time of Gurobi is the time required to guarantee optimality.
}

\begin{tabular}{cccccccccc}
\toprule
\multicolumn{2}{c}{\textbf{Instance}} & \multicolumn{4}{c}{\textbf{\#Optimal}} & \multicolumn{4}{c}{\textbf{Average run time $\mathrm{\, [s]}$}} \\\cmidrule(lr){1-2}\cmidrule(lr){3-6}\cmidrule(lr){7-10}
$n$ & Density $(\%)$ & EIM & QUBO & IHEA\cite{hochbaum2025fast} & Gurobi\cite{hochbaum2025fast} & EIM & QUBO & IHEA\cite{hochbaum2025fast} & Gurobi\cite{hochbaum2025fast} \\
\midrule
\multirow[t]{4}{*}{100} & 25 & \textbf{10} & \textbf{10} & \textbf{10} & \textbf{10} & \textbf{0.0100} & 0.4980 & 1.30 & 0.45 \\
 & 50 & \textbf{10} & \textbf{10} & \textbf{10} & \textbf{10} & \textbf{0.0097} & 0.5551 & 1.29 & 3.71 \\
 & 75 & \textbf{10} & \textbf{10} & \textbf{10} & \textbf{10} & \textbf{0.0171} & 0.6602 & 1.25 & 14.58 \\
 & 100 & \textbf{10} & \textbf{10} & \textbf{10} & \textbf{10} & \textbf{0.0354} & 2.1002 & 1.18 & 25.85 \\
\cmidrule(lr){1-10}
\multirow[t]{4}{*}{200} & 25 & \textbf{10} & 6 & \textbf{10} & \textbf{10} & \textbf{0.0258} & 4.4617 & 2.79 & 7.76 \\
 & 50 & \textbf{10} & 7 & \textbf{10} & \textbf{10} & \textbf{0.8146} & 9.9325 & 3.00 & 50.58 \\
 & 75 & \textbf{10} & 8 & \textbf{10} & 9 & \textbf{0.6401} & 2.8602 & 3.59 & 50.02 \\
 & 100 & \textbf{10} & 5 & \textbf{10} & 8 & \textbf{2.1617} & 3.9242 & 2.83 & 81.99 \\
\cmidrule(lr){1-10}
\multirow[t]{2}{*}{300} & 25 & \textbf{10} & 4 & \textbf{10} & \textbf{10} & \textbf{0.0284} & 29.8289 & 2.93 & 22.36 \\
 & 50 & \textbf{10} & 5 & \textbf{10} & 6 & \textbf{2.9412} & 40.4210 & 3.26 & 91.47 \\
\bottomrule
\end{tabular}

\end{table*}

Table \ref{tab:result} lists the results of solving the QKP. {Concerning the QUBO and EIM, }each solution search was performed with the highest number of iterations, $10^6$, and 10 random seeds, where one iteration is defined as all replicas in RMC completing one MCMC run. {The solution is evaluated at every iteration, and the algorithm stops once the optimal solution is found. The time elapsed until this point is defined as the runtime.}
In the QUBO formulation, the optimal solution was obtained for at least one random seed in every problem involving 100 items. However, for problems with 200 and 300 items, the optimal solutions were only obtained for 35 out of 60 instances. By contrast, the EIM found the optimal solution for all instances, achieving this for all random seeds except for one instance. 
{The EIM also exhibits faster average runtimes than the QUBO formulation.}

For comparison, we show the results of Gurobi\cite{gurobi}, a representative commercial exact solver, and an iterated ``hyperplane exploration" approach (IHEA)\cite{chen2017iterated}, a heuristic solver specialized for QKP, from the evaluation results listed in Ref. \citen{hochbaum2025fast}. IHEA was the fastest among the heuristic solvers in Ref. \citen{hochbaum2025fast} for the problem instances we used in the evaluation. Compared to these methods, EIM obtained optimal solutions for shorter or comparable execution times.

The advantage of the EIM over the QUBO formulation can be attributed to two main factors. First, in EIM, the dependent variables are uniquely determined by the decision variables, meaning that the introduction of such variables does not increase the problem size. Second, in the QUBO formulation, flipping any variable from a feasible solution results in an immediate constraint violation, making exploration more challenging. In contrast, the EIM enables search without constraint violation.

Our experiments showed that the EIM formulation could search for solutions more efficiently than the QUBO formulation. The automatic adjustment of parameters, including the penalty coefficient { and the evaluation of the EIM against existing approaches on more complex instances, remains a challenge for the future.}

\begin{acknowledgment}

We thank Fujitsu Limited for providing us with the extended Ising machine simulator.

\end{acknowledgment}

\bibliographystyle{kudo_short}  
\bibliography{ref} 

\end{document}